\documentstyle[12pt]{article}
\begin{document}
\begin{flushright}
DTP--MSU 95/26 \\ October, 1995
\end{flushright}  

\vskip4cm\begin{center}
{\LARGE\bf
Hidden symmetries in Dilaton--Axion Gravity}\footnote{
Extended version of a talk at the JINR Workshop 
``Geometry and Integrable models'', JINR, Dubna, (Russia), 
October 1994, published in Proceedings,
Published in {\em Geometry and Integrable Models},
P.N.~Pyatov and S.N.~Solodukhin (Eds.), World Scientific, 1996 }\\ \vskip1cm
{\bf
D.V. Gal'tsov} 
and
{\bf  O.V.  Kechkin}\\
\normalsize  Department of Theoretical Physics,  Moscow State University,\\ 
\normalsize Moscow 119899, {\bf Russia}\\
  \vskip2cm
{\bf Abstract}\end{center}
\begin{quote}
Four--dimensional Einstein--Maxwell--dilaton--axion system restricted
to space--times with one non--null Killing symmetry is formulated
as the three--dimensional gravity coupled sigma--model. Several alternative 
representations are discussed and the associated hidden symmetries
are revealed. The action of target space isometries on the initial
set of (non--dualized ) variables is found. New mulicenter
solutions are obtained via generating technique based on the
formulation in terms of the non--dualized variables.
\vskip5mm
\noindent
PASC number(s): 97.60.Lf, 04.60.+n, 11.17.+y
\end{quote}
\newpage

\section{Introduction}
\renewcommand{\theequation}{1.\arabic{equation}}

Einstein equations in General Relativity have a nice property
to admit a three--domensional gravity coupled sigma--model representation,
being restricted to space--times possessing a non--null Killing symmetry.
Moreoer, it turns out that the target space of this sigma--model is 
a symmetric Riemannian space, namely, the coset $SL(2, R)/SO(2)$.
This sigma--model may serve a convenient starting point for a number
of solution--generating methods, most notable being an inverse
scattering transform and Backlund transformations, which arise when
further Killing symmetry commuting with the first one is imposed 
\cite{int}. The existing knowledge of a great variety of analytic
solutions to the Einstein equations is mostly due to this property 
\cite{ex}.

Similar property is shared by other theories
related to General Relativity: Einstein--Maxwell (EM) theory
\cite{ki}, higher dimensional vacuum Einstein equations \cite{ma}, 
bosonic sectors of some dimensionally reduced supergravity theories 
\cite{ju}, \cite{bgm}, \cite{ni}. Crucial feature of all these theories 
is the symmetric space property of the corresponding 
sigma--model in three dimensions (for a review see, e.g., 
Breihtenlohner and Maison \cite{bgm}).
The symmetric space property in its turn requires that the target space
possess a sufficient nubmer of isometries. This convenient way
to investigate the problem is due to the
idea of potential space introduced by Neugebauer and Kramer 
\cite{nk}.

Recent iterest to such models is motivated by the string theory
which is likely to provide a consistent description of gravity at 
the Planck scale.
The most promising string model considered so far is that of 
the heterotic string. In the low--energy (classical) 
limit of this theory with compactified extra dimensions one gets 
in the bosonic sector
the Einstein gravity coupled to massless vector and scalar fields.
 The simplest model of this kind --- 
 ``dilaton--axion gravity'' --- 
incorporates basic features of the full effective action, and can be 
formulated directly in $D=4$ where it includes one $U(1)$ vector 
and two scalar fields coupled in  such a way that the theory possesses
non--abelian $SL(2, R)$ duality 
symmetry. This theory was a subject of recent investigations \cite{gkmult}
which are summarized and developed further xin the present paper.

The Einstein--Maxwell--Dilaton--Axion (EMDA) coupled system contains
a metric $g_{\mu\nu}$, a $U(1)$
vector field $A_\mu$, a Kalb-Ramond antisymmetric tensor field
$B_{\mu\nu}$, and a dilaton $\phi$:
\begin{equation}
S=\frac{1}{16\pi}\int \left\{-R+\frac{1}{3}e^{-4\phi}
H_{\mu\nu\lambda}H^{\mu\nu\lambda}+
2\partial_\mu\phi\partial^\mu\phi -e^{-2\phi}F_{\mu\nu}
F^{\mu\nu}\right\}\sqrt{-g}d^4x,
\end{equation}
where
\[
H_{\mu\nu\lambda}=\partial_{\mu} B_{\nu\lambda} -A_{\mu} F_{\nu\lambda}
\quad + \quad {\rm cyclic},
\]
\begin{equation}
F_{\mu\nu}=\nabla_\mu A_v
-\nabla_\nu A_\mu,
\end{equation}
$\mu ,\nu ,\lambda =0, 1, 2, 3,$ and $\omega_{3L},\; \omega_{3YM}$ are
Lorentz and Yang--Mills Chern--Simons three-forms respectively.
In four dimensions a
Kalb--Ramond field is equivalent to the Peccei--Quinn pseudoscalar axion
$\kappa$,
\begin{equation}
H^{\mu\nu\lambda}=\frac{1}{2}e^{4\phi}E^{\mu\nu\lambda\tau}
\frac{\partial\kappa}{\partial x^\tau},
\end{equation}
and the corresponding Legendre transformation of the action (1.1) reads
\begin{equation}
S=\frac{1}{16\pi}\int \left\{-R+2\partial_\mu\phi\partial^\mu\phi +
\frac{1}{2} e^{4\phi}
{\partial_\mu}\kappa\partial^\mu\kappa
-e^{-2\phi}F_{\mu\nu}F^{\mu\nu}-\kappa F_{\mu\nu}{\tilde F}^{\mu\nu}\right\}
\sqrt{-g}d^4x,
\end{equation}
where ${\tilde F}^{\mu\nu}=\frac{1}{2}E^{\mu\nu\lambda\tau}F_{\lambda\tau}$.

Our purpose is to investigate hidden
symmetries associated with
this model when an additional assumption of stationarity
is imposed on the space--time metric.  
The plan of the papers is as follows. In Sec.\ 2 we outline the derivation of
a three--dimensional sigma--model from the stationary EMDA system.
In Sec.\ 3. complex potentials are introduced which make formulation
similar to the Ernst formulation of vacuum and electrovacuum theories.
The corresponding matrix formulation is given in Sec.\ 4. 
An alternative description in terms of non--dualized quantities
is then suggested (Sec.\ 5) and used to genetrate EMDA multicenter
solutions (Sec.\ 6). 

\section{ $\sigma$--model in three dimensions}
\renewcommand{\theequation}{2.\arabic{equation}}
\setcounter{equation}{0}

The system of equations corresponding to (1.4) consists of the dilaton--axion
modified Maxwell equations
\begin{equation}
\nabla_\nu (e^{-2\phi}F^{\mu\nu}+\kappa{\tilde F}^{\mu\nu})=0,
\end{equation}
 dilaton and  axion equations
\begin{equation}
\nabla_\mu\nabla^\mu\phi =\frac{1}{2}e^{-2\phi}F^2+\frac{1}{2}e^{4\phi}
(\partial\kappa )^2,
\end{equation}
\begin{equation}
\nabla_\mu (e^{4\phi}g^{\mu\nu}\partial_\nu\kappa )+
F_{\mu\nu}{\tilde F}^{\mu\nu}=0,
\end{equation}
and the Einstein equations
\begin{equation}
R_{\mu\nu}=2\phi_{,\mu}\phi_{,\nu}+
\frac{1}{2}e^{4\phi}\kappa_{,\mu}\kappa_{,\nu}+
e^{-2\phi}(2F_{\mu\lambda}{F^\lambda}_\nu +\frac{1}{2}F^2g_{\mu\nu}).
\end{equation}
Here $F^2\equiv F_{\mu\nu}F^{\mu\nu},(\partial\kappa )^2\equiv g^{\mu\nu}
(\partial_\mu\kappa)(\partial_\nu\kappa)$, and $\nabla_\mu$
is the covariant derivative
with respect to the 4--dimensional metric $g_{\mu\nu}$.

Consider a space--time admitting (at least) one Killing vector field which we
choose to be timelike.  Then it is standard to write an interval
using the
three--metric $h_{ij}$, the rotation vector $\omega_i,\; (i, j=1, 2, 3)$
and the scalar $f$, depending on the space coordinates $x^i$, as follows
\begin{equation}
ds^2=g_{\mu\nu}dx^\mu dx^\nu=f(dt-\omega_idx^i)^2-\frac{1}{f}h_{ij}dx^idx^j.
\end{equation}

To derive a three--dimensional $\sigma$--model one has to introduce the
set of appropriate variables. It consists of
the electric $v$ and magnetic $u$ potentials as well as the
twist potential related to $\omega_i$
through a certain curl equation. Electric and magnetic potentials are
introduced through the relations
\begin{equation}
F_{i0}=\frac{1}{\sqrt{2}}\partial_iv,
\end{equation}
\begin{equation}
e^{-2\phi}F^{ij}+\kappa {\tilde F}^{ij}=\frac{f}{\sqrt{2h}}\epsilon^{ijk}
\partial_ku.
\end{equation}
The first substitution solves the spatial part of the Bianchi identity
\begin{equation}
\nabla_\mu{\tilde F}^{\mu\nu}=0,
\end{equation}
while the second satisfies the spatial part of the modified Maxwell equations
(2.1).  From (2.6) and (2.7) the remaining Maxwell tensor components can be
expressed through $v, u, f,\omega_i$ and $h_{ij}$.  Introducing instead of
$\omega_i$ a three vector
\begin{equation}
\tau^i=-f^2\frac{\epsilon^{ijk}}{\sqrt{h}}\partial_j\omega_k,
\end{equation}
one obtains from the remaining Maxwell equations the following set
of three--covariant equations
\begin{equation}
\frac{1}{\sqrt{h}}\partial_i(\frac{1}{f}e^{-2\phi}\sqrt{h}h^{ij}\partial_jv)+
\frac{1}{f^2}\tau^i\partial_iu-
\frac{1}{\sqrt{h}}\partial_i\left[\frac{1}{f}\kappa e^{2\phi}\sqrt{h}h^{ij}
(\partial_ju-\kappa\partial_jv)\right]=0,
\end{equation}
\begin{equation}
\frac{1}{\sqrt{h}}\partial_i\left[\frac{1}{f}e^{2\phi}\sqrt{h}h^{ij}
(\partial_ju-\kappa\partial_jv)\right]-
\frac{1}{f^2}\tau^i\partial_iv=0.
\end{equation}
The equations for dilaton and axion fields (2.2), (2.3) with account for
(2.5) will take the form
\begin{equation}
\frac{1}{\sqrt{h}}\partial_i(\sqrt{h}h^{ij}\partial_j\phi )=\frac{1}{2f}
\left[e^{-2\phi}\partial_iv\partial^iv-e^{2\phi}
(\partial_iu-\kappa\partial_iv)
(\partial^iu-\kappa\partial^iv)\right]
+\frac{e^{4\phi}}{2}\partial_i\kappa\partial^i\kappa ,
\end{equation}
\begin{equation}
\frac{1}{\sqrt{h}}\partial_i(\sqrt{h}h^{ij}e^{4\phi}\partial_i\kappa )=
\frac{2}{f}e^{2\phi}(\partial_iu-\kappa\partial_iv)\partial^iv,
\end{equation}
where (and in what follows) raising and lowering of the three--indices
is understood with respect to the three--metric $h_{ij}$ and its inverse
$h^{ij}$.

Remarkably, the mixed components of the Einstein equations (2.4)
remain unaffected by neither a dilaton nor an axion:
\begin{equation}
\frac{1}{f}R_0^i\equiv -\frac{1}{2\sqrt{h}}\epsilon^{ijk}\partial_k{\tau_j}=
\frac{\epsilon^{ijk}}{\sqrt{h}}\partial_jv\partial_ka.
\end{equation}
This equation can be integrated giving the twist potential $\chi$
\begin{equation}
\tau_i=\partial_i\chi +v\partial_iu-u\partial_iv.
\end{equation}
An equation for $\chi$ can be found by taking a three--covariant divergence
of (2.15) multiplied by $f^{-2}$ with account for the definition (2.9):
\begin{equation}
f(\Delta\chi +v\Delta u-u\Delta v)=2(\partial_i\chi +v\partial_i u-u
\partial_iv)\partial^if,
\end{equation}
where
\begin{equation}
\Delta =\frac{1}{\sqrt{h}}\partial_i(\sqrt{h}h^{ij}\partial_j)
\end{equation}
is a three--dimensional Laplacian.  Finally, an equation for $f$
follows from the
(00)--component of the Einstein equations in a similar form
\begin{equation}
f\Delta f-\partial_if\partial^if+\tau_i\tau^i=
f\left[e^{2\phi}(\partial_iu-\kappa\partial_iv)(\partial^iu-
\kappa\partial^iv)+e^{-2\phi}\partial_iv\partial^iv\right]
\end{equation}
where the definition (2.15) has to be used.

The system of 6 equations (2.10)--(2.13), (2.16), (2.18) constitutes
the set of equations for a curved space $\sigma$--model  with the
6--dimensional target space.  The remaining (spatial) Einstein equations
from (2.4) may be consistently regarded as 3--dimensional Einstein equations
for the metric $h_{ij}$.  Their source part
is derivable from the same sigma--model as the 3--dimensional
energy momentum tensor
\[
{\cal R}_{ij}=\frac{1}{2f^2}(f_{,i}f_{,j}+\tau_i\tau_j)+2\phi_{,i}
\phi_{j}+\frac{1}{2}e^{4\phi}\kappa_{,i}\kappa_{,j}-
\]
\begin{equation}
-\frac{1}{f}\left[e^{-2\phi}v_{,i}v_{,j}+
e^{2\phi}(u_{,i}-\kappa v_{,i})(u_{,j}-\kappa v_{,j})\right],
\end{equation}
where $\tau_i$ is given by (2.15), and ${\cal R}_{ij}$ is the
three--dimensional Ricci tensor.

Now, it can be easily checked that all 3--dimensional equations may be
obtained by a variation of the action
\[
S=\int\Bigl\{{\cal R}-\frac{1}{2f^2}[(\nabla f)^2+(\nabla\chi +
v\nabla u-u\nabla v)^2]-2(\nabla\Phi )^2-
\]
\begin{equation}
-\frac{1}{2}e^{4\phi}(\nabla\kappa )^2+\frac{1}{f}\left[e^{2\phi}
(\nabla u-\kappa\nabla v)^2+
e^{-2\phi}(\nabla v)^2\right]\Bigr\}\sqrt{h}d^3x,
\end{equation}
where ${\cal R}\equiv {\cal R}_i^i$ and $\bf\nabla$ stands for 3--dimensional
covariant derivative.  

This action  can be rewritten as the gravity coupled three--dimensional
$\sigma$--model 
\begin{equation}
S=\int \left({\cal R}-{\cal G}_{AB}\partial_i\varphi^A\partial_j\varphi^B
h^{ij}\right)\sqrt{h}d^3x,
\end{equation}
where $\varphi^A=(f,\; \chi,\; v,\;u,\;\kappa,\; \phi,)\; A=1,..., 6$.
The corresponding target space metric reads
\begin{eqnarray}
{dl}^2 & = & \frac{1}{2}f^{-2}[{df}^2+
(d\chi +vdu-udv)^2]-f^{-1}[e^{2\phi }{(du-\kappa dv)}^2
+e^{-2\phi }{dv}^2] \nonumber \\
& + & 2{d\phi }^2+
\frac{1}{2}e^{4\phi }{d\kappa }^2.
\end{eqnarray}
This expression generalizes Neugebauer and Kramer potential space metric 
found for the stationary Einstein--Maxwell system in 1969 \cite{nk}.
It is worth noting that the present $\sigma$--model does not reduce to the
Einstein--Maxwell one if $\kappa=\phi=0$, because the equations for
a dilaton  and an axion  generate constraints $F^2=F\tilde F =0$. 
Hence, generally the solutions
of the Einstein--Maxwell theory with one Killing symmetry are not related
to solutions of the present theory by target space isometries (except for
the case $F^2=F\tilde F =0$). From the other hand, one can consistently set
in the present $\sigma$--model $f=\chi=\phi=\kappa=0$ reducing it to the Einstein
vacuum sigma--model. Therefore all solutions to the vacuum Einstein
equations with one non--null Killing symmetry 
are related to some solutions of the
system in question by target space isometries. 

\section{Complex potentials}
\renewcommand{\theequation}{3.\arabic{equation}}
\setcounter{equation}{0}

More concise formulation of the theory can be achieved in therms 
of the following complex variables
\begin{equation}
z = \kappa + i e^{-2\phi}
\end{equation}
\begin{equation}
\Phi = u - zv
\end{equation}
\begin{equation}
E = if - \chi + v\Phi
\end{equation}
The first is the standard complex dilaton--axion field, while two other
may be regarded as suitable generalizations of the Ernst potentials.
In terms of them the target space metric reads
\begin{equation}
dl^2=\frac{1}{2f^2}\left|dE+\frac{2{\rm Im}\Phi}{{\rm Im}z} d\Phi-
\left(\frac{{\rm Im}\Phi}{{\rm Im}z}\right)^2dz\right|^2-
\frac{1}{f{\rm Im}z}\left|d\Phi-\frac{{\rm Im}\Phi}{{\rm Im}z}dz \right|^2+
\frac{|dz|^2}{2({\rm Im}z)^2}.
\end{equation}

It can be shown that this three--dimensional complex manifold is K\"ahler. 
Indeed, in terms of the complex coordinatess 
$z^\alpha=(E, -z, \Phi),\;z^{\bar\alpha}=({\bar E }, {-\bar z}, 
{\bar \Phi})$, 
where $z^{\bar\alpha}$ is a shorthand for ${\bar z^{\bar \alpha}}$, one
can write an Hermitian metric on the target space as
\begin{equation}
dl^2=K_{\alpha{\bar\beta}}dz^{\alpha} dz^{\bar\beta}.
\end{equation}
This implies the existence of a non--degenerate two--form
\begin{equation}
\Omega=\frac{i}{2}K_{\alpha{\bar\beta}}dz^{\alpha} \wedge dz^{\bar\beta}.
\end{equation}
One can easily check that this form is closed:
\begin{equation}
d\Omega=0,
\end{equation}
what means that we have a K\"ahler manifold.

Two describe isometries of the target space it is convenient to
take into account the following two dicrete symmetries. The first
is quite obvious
\begin{equation}
z^{'}=E,\quad  E^{'}=z,\quad \Phi^{'}=\Phi.  
\end{equation}
Another useful discrete symmetry operation, which will be called 
``double prime'', is 
\begin{equation}
E^{''} = \frac{z}{Ez+\Phi^2},\quad
z^{''} = \frac{E}{Ez+\Phi^2},\quad
\Phi^{''} =\frac{\Phi}{Ez+\Phi^2}.
\end{equation}
 
The set of coninuous isometries consists of three
gauge (electric $e$, magnetic $m$, and gravitational $g$) transformations,
a scale transformation ($s$), $SL(2, R)$ duality ($d_1, d_2, d_3$),
two Harrison--type ($H_1, H_2$) and Ehlers--type ($E$) transformations.
The Killing
vectors corresponding to gauge transformations 
in terms of complex potentials read
\begin{equation}
K_e=2\Phi\partial_E - z\partial_\Phi +c.c.,
\end{equation}
\begin{equation}
K_m=\partial_\Phi +c.c.,
\end{equation}
\begin{equation}
K_g=\partial_E +c.c..
\end{equation}
The scale generator is
\begin{equation}
K_s=2E\partial_E+ \Phi\partial_\Phi +c.c..
\end{equation}

The Ehlers--Harrison--type generators 
\cite{ki} can be obtained by ``priming'' the gauge generators
\begin{equation}
K_{H_1}= K_e^{'}=2\Phi\partial_z - E\partial_\Phi +c.c.,
\end{equation}
\begin{equation}
K_{H_2}= K^{''}_m=(Ez-\Phi^2)\partial_\Phi- 2\Phi(z\partial_z +
E\partial_E) +c.c.,
\end{equation}
\begin{equation}
K_{E}= K^{''}_g=\Phi^2\partial_z-E(E\partial_E+\Phi\partial_\Phi) +c.c.
\end{equation}

Two generators of the $S$--duality subgroup correspond to the primed 
$g$ and $s$ transformations
\begin{equation}
K_{d_1}= K_g^{'}=\partial_z +c.c.,
\end{equation}
\begin{equation}
K_{d_3}= K_s^{'}=2z\partial_z +\Phi\partial_\Phi + c.c.,
\end{equation}
while the remaining one may be obtained by priming the Ehlers generator
\begin{equation}
K_{d_2}= K^{'}_E=\Phi^2\partial_E-z(z\partial_z+\Phi\partial_\Phi) +c.c.
\end{equation}

It can be checked that these 10 Killing vectors form the algebra $sp(4,R)$,
while the target space in isomorphic to the coset $Sp(4,R)/U(2)$.
Finite $Sp(4, R)$ transformations may also be obtained starting with 
the gravitational 
\begin{equation}
E=E_0+\lambda,\quad \Phi=\Phi_0,\quad z=z_0,
\end{equation}
electric
\begin{equation}
E=E_0-2\lambda \Phi_0-\lambda^2 z_0,\quad 
\Phi=\Phi_0+\lambda z_0,\quad z=z_0,
\end{equation}
the magnetic
\begin{equation}
E=E_0,\quad \Phi=\Phi_0+\lambda,\quad z=z_0,
\end{equation}
gauge as well as the scale transformation
\begin{equation}
E=e^{2\lambda}E_0,\quad \Phi=e^{\lambda}\Phi_0,\quad z=z_0,
\end{equation}
(which can be easily integrated from the infinitesimal form).
The remaining six group elements can be found using prime
operations. The $Sl(2, R)$--duality subgroup read
\begin{eqnarray}
&& d_1:\qquad E=E_0,\quad \Phi=\Phi_0,\quad z=z_0+\lambda,\nonumber\\
&& d_2:\qquad E=E_0+\lambda\frac{\Phi_0^2}{1+\lambda z_0},\quad 
\Phi=\frac{\Phi_0}{1+\lambda z_0},\quad 
z=\frac{z_0}{1+\lambda z_0},\\ 
&& d_3:\qquad E=E_0,\quad \Phi=e^{\lambda}\Phi_0,\quad 
z=e^{2\lambda}z_0,\nonumber 
\end{eqnarray}
the electric and magnetic Harrison transformations are
\begin{equation}
E=E_0,\quad \Phi=\Phi_0+\lambda E_0,\quad 
z=z_0-2\lambda \Phi_0-\lambda^2 E_0,
\end{equation}
\begin{equation}
E=\frac{E_0}{(1+\lambda\Phi_0)^2+\lambda^2 E_0 z_0},\quad
\Phi=\frac{\Phi_0(1+\lambda\Phi_0)+\lambda E_0 z_0}
{(1+\lambda\Phi_0)^2+\lambda^2 E_0 z_0},\quad
z=\frac{z_0}{(1+\lambda\Phi_0)^2+\lambda^2 E_0 z_0},
\end{equation}
while the Ehlers--type transformation reads
\begin{equation}
E=\frac{E_0}{1+\lambda E_0},\quad
\Phi=\frac{\Phi_0}{1+\lambda E_0},\quad 
z=z_0+\frac{\lambda\Phi_0^2}{1+\lambda E_0}. 
\end{equation}

The isometry group acts transitively on the target space which is
a symmetric Riemannian space. The direct way to show this is 
to introduce matrix representation in terms of symmetric symplectic
$4\times 4$ matrices forming a coset $Sp(4, R)/U(2)$.

\section{Matrix representation}
\renewcommand{\theequation}{4.\arabic{equation}}
\setcounter{equation}{0}
 
For any matrix $G\in Sp(4, R)$ one can perform a Gauss decomposition
\begin{equation}
G=G_LG_SG_R,
\end{equation}
where
\begin{equation}
G_R=\left(\begin{array}{crc}
I&R\\
O&I\\
\end{array}\right),\quad
G_S=
\left(\begin{array}{crc}
{S^T}^{-1}&O\\
O&S\\
\end{array}\right),\quad
G_L=
\left(\begin{array}{crc}
I&O\\
L&I\\
\end{array}\right),
\end{equation}
and $S, R, L$ are real  $ 2\times 2 $ matrices, $R, L$ being symmetric,
$R^T=R,\;L^T=L$. 
Useful relations are
\begin{equation}
G_{R_1}G_{R_2}=G_{R_1+R_2},\quad G_{S_1}G_{S_2}=G_{S_1S_2},\quad 
G_{L_1}G_{L_2}=G_{L_1+L_2} ,
\end{equation}
so that for inverse matrices one has 
$G_R^{-1}=G_{-R},\;\;G_S^{-1}=S_{-R},\;\;G_L^{-1}=G_{-L}$.
With this parametrization 
\begin{equation}
G=\left(\begin{array}{crc}
{S^T}^{-1}&{S^T}^{-1}R\\
L{S^T}^{-1}&S+L{S^T}^{-1}R\\
\end{array}\right).
\end{equation}

A subset of symmetric symplectic matrices $M\in Sp(4,R),\;M^T=M$ 
represents a coset
$Sp(4, R)/U(2)$. The corresponding Gauss decomposition reads
\begin{equation}
M={\cal Q}^T{\cal P}{\cal Q},
\end{equation}
 \begin{equation}
{\cal Q}=
\left(\begin{array}{crc}
I&Q\\
O&I\\
\end{array}\right),\quad
{\cal P}=
\left(\begin{array}{crc}
P^{-1}&O\\
O&P\\
\end{array}\right),
\end{equation}
where two real symmetric $2\times 2$ matrices are introduced
$ Q^T=Q,\; P^T=P$. Multiplying  matrices we obtain
\begin{equation} 
M=\left(\begin{array}{crc}
P^{-1}&P^{-1}Q\\
QP^{-1}&P+QP^{-1}Q\\
\end{array}\right).
\end{equation}

It is useful to combine $P$ and $Q$ into one complex symmetric matrix
\begin{equation}
Z=Q+iP,
\end{equation}
which transforms under an action of  $G_R, G_L, G_S $ as follows:
\begin{equation}
Z \rightarrow Z+R,
\end{equation}
for $G=G_L$ :
\begin{equation}
Z^{-1} \rightarrow Z^{-1}-L,
\end{equation}
 while for $G=G_S$ :
\begin{equation}
Z \rightarrow S^TZS.
\end{equation}
Here the following transformation law for $M$ is assumed
\begin{equation}
M\rightarrow G^TMG.
\end{equation}

These transformations are precisely the matrix analog of the
$SL(2, R)$ transformations which hold for the usual scalar 
dilaton--axion field (3.1)
under S--duality generated by $K_{d_1},\;K_{d_2},\;K_{d_3}$.

It can be shown that the explicit expression for $Z$ in terms
of the complex  potentials introduced in the previous section is very simple
\begin{equation}
Z=\left(\begin{array}{crc}
E&\Phi\\
\Phi&-z\\
\end{array}\right).
\end{equation}
 
Sigma--model of the Sec. 3 can now be rewritten as a coset model for the 
matrix $M$ (4.7). Let us introduce a matrix current
\begin{equation}
{ J^M}=\nabla  MM^{-1}.
\end{equation}
Using (4.8) into we get
{\small
\[ J^M=
\left(\begin{array}{rrr}
-P^{-1}\nabla P-P^{-1}\nabla QP^{-1}Q & \hspace{2mm} &
P^{-1}\nabla QP^{-1}\\
\nabla Q-QP^{-1}\nabla P-\nabla PP^{-1}Q-QP^{-1}\nabla QP^{-1}Q &\hspace{2mm} &
\nabla PP^{-1}+QP^{-1}\nabla QP^{-1}
\end{array}\right).\]
}
In this notation, the three--dimensional sigma--model action 
can now be written as
\begin{equation}
 S=\int \left(-{\cal R}+\frac{1}{4}Tr({J^M}^2)\right)\sqrt{h}d^3x,
\end{equation}
while the corresponding field equations read
\begin{equation}
\nabla {  J^M}=0,
\end{equation}
\begin{equation}
{\cal R}_{mn}=\frac{1}{4}Tr({ J^M}_m{ J^M}_n).
\end{equation}

 There exists also a concise representation of the same model
directly by $2\times 2$ matrices.
Indeed, from three $2\times 2$ matrix equations following the present
representation
\begin{equation}
\nabla \left(P^{-1}\nabla Q P^{-1}\right)=0,
\end{equation}
\begin{equation}
\nabla \left(QP^{-1}\nabla Q P^{-1}+\nabla P P^{-1}\right)=0 , 
\end{equation}
\begin{equation}
\nabla \left(\nabla Q -\nabla P P^{-1} Q- Q P^{-1}\nabla P
-QP^{-1}\nabla Q P^{-1} Q\right)=0 , 
\end{equation}
only first two are independent. The last can be rewritten using first two 
as
\begin{equation}
P\nabla \left(P^{-1}\nabla Q P^{-1}\right)P=0.
\end{equation}
 Introducing two $2\times 2$ matrix currents
\begin{equation} 
{J}_1=\nabla \it PP^{-1},\quad
{ J}_2=\nabla \it QP^{-1}.
\end{equation}
we can rewrite the equations of motion as
\[
\nabla { J}_1+{{ J}_2}^2=0,
 \] 
\begin{equation}
\nabla {J}_2-{J}_1{ J}_2=0,
\end{equation}
 \[
{\cal R}_{mn}=\frac{1}{2}Tr\left({ J}_{1m}{ J}_{1n}+
{ J}_{2m}{J}_{2n}\right).
 \]                                     
It can also be checked that
\begin{equation}
Tr \left( { J}_1^2+{ J}_2^2\right)=\frac{1}{2}Tr{ J}^2 .
\end{equation}

Further simplification comes from the introduction of a complex
matrix dilaton--axion $Z$ instead of its real and imaginary parts
(from an explicit for of $Z$ ) it is clear that real and imaginary parts 
are more involved because of the complicated structure of the denominator).
Then instead of two real $2\times 2$ currents one can build
one complex matrix current
\begin{equation}
{ J}_Z=\nabla Z{(Z-\overline{Z})}^{-1},
\end{equation}
and cast the action into the following form
\begin{equation}
S=\int\frac{1}{2}\left(-{\cal R}+
2Tr{J}_Z\overline{{ J}_Z}\right)\sqrt{h}d^3x .
\end{equation}
The correspoding equations read
\[
\nabla {J}_Z={ J}_Z({ J}_Z+{\bar{ J}_Z}),
\]
\begin{equation}
{\cal R}_{mn}=2Tr{J}_{Zm}{\bar{ J}_{Zn}}.
\end{equation}                                       
 
The metric of the target space can be presented 
in terms of the matrix--valued
one--forms corresponding to the above currents
\begin{equation}
\omega=dMM^{-1},\quad \omega_1=dPP^{-1},\quad \omega_2=dQP^{-1},\quad
\omega_Z=dZ (Z-{\bar Z})^{-1},\quad
\end{equation}
by three alternative ways
\begin{equation}
{ds}^2=\frac{1}{4}Tr(\omega^2)= \frac{1}{2}Tr(\omega_1^2+\omega_2^2)=
2 Tr(\omega_Z {\bar \omega_Z}).
\end{equation}

\section{Back to non--dualized variables }
\renewcommand{\theequation}{5.\arabic{equation}}
\setcounter{equation}{0}

To achieve sigma--model description of the system in three dimensions 
it was essential to use the dualized variables for the rotational
part of a metric and for the magnetic part of a vector field, 
as well as to introduce the Peccei--Quinn pseudoscalar counterpart to the
antisymmetric Kalb--Ramond field. This description was particularly
useful in deriving hidden symmetries of the dimensionally reduced 
system. Once they are found, one can go back to the initial formulation
and to reveal their action on the non--dualized set of variables.
Apart from purely formal interest, this also opens a way to obtain
an infinite symmetry algebra in the further two--dimensional reduction.

Transition to the initial variables may now be regarded as the 
formal solution of the matrix equation
\begin{equation}
\nabla J^{M} = 0,\quad J^{M}=\nabla M M^{-1}.
\end{equation}
Since its divergence is zero, the matrix current can be presented
as a curl of some vector matrix $\vec N$ 
\begin{equation}
\nabla \times \vec N =J^{M}.
\end{equation}
It can be checked that an alternative form of the equation
$\nabla J^{M} = 0$ reads
\begin{equation}
\nabla \times J^{M}-J^{M}\times J^{M} =0.
\end{equation}

In terms of $2\times 2$ symmetric matrices
$P$ ( $Q$ one gets the following set of equations
\begin{equation}
\nabla \times \vec \Omega =P^{-1}\nabla Q P^{-1},
\end{equation}
\begin{equation}
\nabla \times \vec U =\nabla PP^{-1} + QP^{-1}\nabla QP^{-1},
\end{equation}
\begin{equation}
\nabla \times \vec V =\nabla Q - \nabla PP^{-1}Q-QP^{-1}\nabla P -
QP^{-1}\nabla QP^{-1}Q,
\end{equation}
where $\vec \Omega $, $\vec V$ and $\vec U$ are symmetric real $2\times 2$ 
matrices forming $\vec N$ as follows
\begin{equation}
{\vec N}=
\left( \begin{array}{ccc}
-\vec U^T & \vec \Omega \\
\vec V & \vec U \\
\end{array} \right).
\end{equation}

From explicit expressions for $P$ and $Q$
\begin{displaymath}
{P}=
\left( \begin{array}{ccc}
f-e^{-2\phi}v^2 & -e^{-2\phi}v\\
-e^{-2\phi}v  & -e^{-2\phi}\\
\end{array} \right),
\end{displaymath}
\begin{displaymath}
{Q}=
\left( \begin{array}{ccc}
wv-\chi & w\\
w  & -\kappa\\
\end{array} \right)
\end{displaymath}
it is clear that they provide a natural decomposition of the full 
set of sigma--model coordinates into non--dualized ($f, \phi, v$) and
dualized ($\chi, u, \kappa$) subsets. Hence, it can be anticipated
that the matrix $P$ will remain unchanged, while $Q$ will be replaced by
$\vec \Omega $. After some calculations one can express it through
the non--dualized variables as follows: 
\begin{equation}
{\vec \Omega}=
\left( \begin{array}{ccc}
\vec \omega & -(\vec A+A_0\omega )\\
-(\vec A+A_0\omega )& -\vec B+A_0(\vec A+A_0\omega )\\
\end{array} \right),
\end{equation}
where $ A_0=v, B_i=B_{0i} $,
$ \omega _i$ is the metric function and $ A_i $ is three--dimensional 
vector--potential defined as 
$\sqrt{2}F_{ij}=(\partial_i A_j -\partial_j A_i )$. 
Two other matrices $\vec U$ and $\vec V$ can be found once
$\vec \Omega $ is known (all three--dimensional vector operations 
in the above formulas have to be effected using the metric $h^{ij}$. 

It is worth noting that the spatial components of the Kalb--Ramond
field do not enter into the above set of variables. In fact, 
in the stationary case they can be expressed  through other variables. 
Introducing a three--dimensional dual to $B_{ij}$
\begin{equation}
C^i=\frac {1}{2}E^{ijk}B_{jk},
\end{equation}
from stationarity $ \kappa _{,0} =0 $ one can find after some
rearrangement
\begin{equation}
\nabla \vec C=\nabla (\vec B\times \vec \omega)+[\vec B-A_0(\vec A+A_0\vec
\omega)]\nabla \times \vec \omega+(\vec A+A_0\vec \omega)\nabla \times
(\vec A+A_0\vec \omega).
\end{equation}

One can derive the three--dimensional action variation of which
gives the equations for matrices $P$ and $\vec \Omega$. First,
from the Eq.(5.4) one gets
\begin{equation}
\nabla \times [P(\nabla \times \vec \Omega) P]=0,
\end{equation}
while (5.5) takes the form
\begin{equation}
\nabla (\nabla PP^{-1})+P(\nabla \times \vec \Omega) 
P\nabla \times \vec \Omega=0.
\end{equation}
Together with the Einstein equations
\begin{equation}
R_{ij}=2Tr(J^P_iJ^P_j+J^{\Omega}_iJ^{\Omega}_j),
\end{equation}
where
\begin{equation}
J^P=\nabla PP^{-1},\qquad J^{\Omega}=P\nabla \times \vec \Omega
\end{equation}
we have the complete description of the system in terms of
 $P$ and $\vec \Omega$. The Eqs. (5.11-12) can be rewritten as
\begin{equation}
\nabla J^P + (J^{\Omega})^2=0,\qquad
\nabla \times J^{\Omega}-J^{\Omega}\times J^P=0,
\end{equation}
while the corresponding action read
\begin{equation}
S=\int d^3xh^{\frac {1}{2}}(-R+2Tr[(J^P)^2-(J^{\Omega})^2]).
\end{equation}

The equations for $\vec U$ and $\vec V$ still contain undesired
matrix $Q$. However two ``Legendre transformed'' quantities
\begin{equation}
\vec U^{'}=\vec U-Q\vec \Omega, \quad
 \vec V^{'}=\vec V + Q\vec \Omega Q +\vec U^{'}Q+Q\vec U^{'T}
\end{equation}
satisfy the equations depending only on $P$ ( $\vec \Omega$:
\begin{equation}
\nabla \times \vec U^{'} =\nabla PP^{-1}-P(\nabla \times \vec \Omega) P
\times \vec \Omega
\end{equation}
\begin{equation}
\nabla \times \vec V^{'}=P(\nabla \times \vec \Omega) P+
P(\nabla \times \vec \Omega) P\times \vec U^{'T}-
\vec U^{'}\times P(\nabla \vec \Omega) P.
\end{equation}
Similarly, a non--dynamical variable $\vec C$ may be replaced by
\begin{equation}
\vec C^{'}=\vec C-\frac {1}{2}[\vec B+
A_0(\vec A+A_0\vec \omega )]\times\vec \omega,
\end{equation}
the equation for which contains only ``allowed'' variables:
\begin{equation}
\nabla \vec C^{'}=\frac {1}{2}Tr(\vec \Omega \epsilon \nabla \times
\vec \Omega \epsilon).
\end{equation}
where
\begin{displaymath}
{\bf \epsilon}=
\left(\begin{array}{ccc}
0&1\\
-1&0
\end{array} \right).
\end{displaymath}

Target space isometries induce certain transformations of the 
matrix $\vec \Omega$. These can be found by a direct computation
using the Eqs. (5.4--6). The subgroup $R$ leaves
$\nabla P$ and $\nabla \times \vec \Omega$ unchanged, and now some
more general transformation of the same kind can be foun
\begin{equation}
P\rightarrow P,\qquad \vec \Omega \rightarrow \vec \Omega +\nabla R^{'},
\end{equation}
where $R^{'}$ is an arbitrary {\em non--constant} symmetric matrix.

Now, for the $S$--transformation it is convenient to redefine
\begin{equation}
S^{'}=S^{-1T},
\end{equation}
so that we get the following counterpart 
\begin{equation}
P^{-1}\rightarrow S^{'T}P^{-1}S^{'},
\end{equation}
\begin{equation}
\vec \Omega \rightarrow S^{'T}\vec \Omega S^{'}.
\end{equation}

Finally, for the $L$--transformations one has
\begin{equation}
P^{-1}=P_0^{-1}+LQ_0P_0^{-1}+P_0^{-1}Q_0L+L(P_0+Q_0P_0^{-1}Q_0)L,
\end{equation}
while $\vec \Omega$ can be shown to transform as
\begin{equation}
\vec \Omega \rightarrow \vec \Omega +L\vec U+\vec U^TL+L\vec V L.
\end{equation}
 
For generating purposes it is convenient to combine the action of
$S, L$ transformations. Omitting primes, one gets
\begin{equation}
P^{-1}=S^T[P_0^{-1}+LQ_0P_0^{-1}+P_0^{-1}Q_0L+L(P_0+Q_0P_0^{-1}Q_0)L]S
\end{equation}
\begin{equation}
\vec \Omega =S^T[\vec \Omega _0+L\vec U_0+\vec U_0^TL-L\vec V_0L]S.
\end{equation}

It is worth noting that for the three--dimensional description in terms
of the non--dualized quantities we have introduced more variables than 
in the previous sigma--model formulation. The excessive variables
could in principle be eliminated by fixing the gauge, but then 
three--dimensional covariance would be lost.

\section{Multicenter solutions}
\renewcommand{\theequation}{6.\arabic{equation}}
\setcounter{equation}{0}
Here we give some examples of the application of symmetries to derive
multicenter solutions to dilaton--axion gravity. Le us specify the 
transformation formulas for asymptotically flat solutions with zero
asymptotic values of material fields. In this case
\begin{equation}
P_{0\infty}^{-1}=\sigma _3,\qquad Q_{0\infty}=\vec \Omega _{0\infty}=0.
\end{equation}
Demanding that the $L$--transformation leave unchanged asymptotic
values of $f,\phi ,v$, one obtains for the matrix $L$ the following
expression
\begin{equation}
L=l\Sigma ,
\end{equation}
where
\begin{displaymath}
{\Sigma}=
\left(\begin{array}{ccc}
1&\sigma \\
\sigma &1
\end{array} \right).
\end{displaymath}
and $\sigma =\pm 1$,  $l=const$. We will assume that the
$S$--transformation leads to a non--trivial asymptotic for the dilaton 
\begin{displaymath}
{P_{\infty}^{-1}}=
\left(\begin{array}{ccc}
1&0\\
0&-e^{2\phi _{\infty}}
\end{array} \right).
\end{displaymath}
These assumptions leave one free parameter in $S$: 
\begin{displaymath}
{S}=
\left(\begin{array}{ccc}
ch\theta&e^{\phi _{\infty}}sh\theta\\
sh\theta&e^{\phi _{\infty}}ch\theta
\end{array} \right).
\end{displaymath}
As a whole we get two parameter ($l$ and $\theta$) family of solutions
applying $S, L$--transformation to a chosen asumptotically flat seed 
solution.

As a first example we take as a seed
\begin{equation}
P_0^{-1}=\sigma _3, \qquad \vec \Omega _0=\vec \Lambda \Sigma {'}
\end{equation}
where $\Sigma {'}$ corresponds to the replacement of $\sigma $ to
$\sigma {'}$ in (). A (non--constant) matrix vector $\vec \Lambda$ is
introduced according to
\begin{equation}
\nabla \times \vec \Lambda =\nabla \lambda, \qquad
 {\bf \lambda}=
\sum_{i=1}^{N}\frac {\lambda _i}{\vert \vec r-\vec r_i\vert},
\end{equation}
where $\vec r_i$  are positions of the sources.
Then three--metric turns out to be flat 
\begin{equation}
dl^2=d\vec r^2.
\end{equation}
Physically this solution corresponds to an equilibrium  system of an 
arbitrary number of point--like magnetic monopoles endowed with axion 
charges (magnetic repulsion is balanced by axion attraction). This field
configuration generates twist, so that the resuling metric is a massless
multi--Taub--NUT. 

Now we apply to this solution a combined $S, L$ transformation described
above () setting $\sigma ^{'}=-\sigma $ (only in this case 
$L$--transformation gives non--trivial result). One gets
\begin{displaymath}
{P^{-1}}=(1+4l\lambda)
\left(\begin{array}{ccc}
1&0\\
0&-e^{2\phi _{\infty}}
\end{array} \right),
\end{displaymath}
\begin{displaymath}
{\vec \Omega}=\vec \Lambda \left (e^{-2\sigma \theta}
\left(\begin{array}{ccc}
1&-\sigma e^{\phi _{\infty}} \\
-\sigma e^{\phi _{\infty}}&e^{2\phi _{\infty}}
\end{array} \right)
-4l^2e^{2\sigma \theta}
\left(\begin{array}{ccc}
1&\sigma e^{\phi _{\infty}} \\
\sigma e^{\phi _{\infty}}&e^{2\phi _{\infty}}
\end{array} \right)\right ).
\end{displaymath}
In physical terms
\begin{equation}
f^{-1}=1+4l\lambda,\qquad e^{2\phi}=e^{2\phi _{\infty}}f^{-1},
\qquad A_0=0
\end{equation}
\begin{equation}
\vec \omega =(e^{-2\sigma \theta}-4l^2e^{2\sigma \theta})\vec \Lambda
\qquad
\vec A=\sigma e^{\phi _{\infty}}
(e^{-2\sigma \theta}+4l^2e^{2\sigma \theta})\vec \Lambda,\qquad
\vec B=-e^{2\phi _{\infty}}\vec \omega.
\end{equation}
In the case
\begin{equation}
e^{-4\sigma \theta}=4l^2,
\end{equation}
after redefinitions 
$2l\lambda \rightarrow \lambda, \sigma l/|l|\rightarrow \sigma$ 
the solution can be written as 
 \begin{equation}
f^{-1}=e^{2(\phi -\phi _{\infty})}=1+2\lambda,\qquad \vec A=2\sigma
e^{\phi _{\infty}}\vec \Lambda,\qquad A_0=\vec \omega=\vec B=0.
\end{equation}
and therefore describes the system of point massive magnetic charges
endowed with dilaton charges. In this case (no axion field) twist
is not generated.

As a second example consider an $(S,L)$--transformation of the
electric Majumdar--Papapetrou--like solution
\begin{equation}
P_0=\sigma _3+\lambda \Sigma,\qquad \vec \Omega _0=0,
\qquad dl^2=d\vec r^2,
\end{equation}
corresponding to 
\begin{equation}
Q_0=\vec \Omega _0=\vec V_0=0,
\end{equation}
\begin{displaymath}
{\vec U_0}=\vec \Lambda
\left(\begin{array}{ccc}
1&-\sigma \\
\sigma &-1
\end{array} \right),
\end{displaymath}.
With no additional assumptions for 
$\sigma ^{'}$ ( $\sigma$, one obtains
{\normalsize
\begin{displaymath}
{P^{-1}}=
\left(\begin{array}{ccc}
1&0 \\
0&e^{-2\phi _{\infty}}
\end{array} \right)
- \lambda \left (e^{-2\sigma \theta}
\left(\begin{array}{ccc}
1&-\sigma e^{\phi _{\infty}} \\
-\sigma e^{\phi _{\infty}}&e^{2\phi _{\infty}}
\end{array} \right)
-2l^2(1+\sigma \sigma ^{'})e^{2\sigma \theta}
\left(\begin{array}{ccc}
1&\sigma e^{\phi _{\infty}} \\
\sigma e^{\phi _{\infty}}&e^{2\phi _{\infty}}
\end{array} \right)\right ),
\end{displaymath}
}
\begin{displaymath}
{\vec \Omega}=2l\vec \Lambda \left (
\left(\begin{array}{ccc}
1&0 \\
0&-e^{2\phi _{\infty}}
\end{array} \right)
+\sigma \sigma ^{'}
\left(\begin{array}{ccc}
ch2\theta &e^{\phi _{\infty}}sh2\theta \\
e^{\phi _{\infty}}sh2\theta &e^{2\phi _{\infty}}ch2\theta
\end{array} \right)\right ).
\end{displaymath}
Extracting physical parameters, one can see that we have generated
magnetic, axion and NUT charges.

\end{document}